\documentclass[12pt,a4paper]{article}
\usepackage{epsf,array,delarray,amsmath,stmaryrd,amssymb}
\usepackage{graphicx}
\usepackage{float}

\author{ L.~C.~G.  Rogers\\
Statistical Laboratory, University of Cambridge
\footnote{
Statistical Laboratory, University of Cambridge, 
Wilberforce Road, Cambridge CB3 0WB, UK;
 L.C.G.Rogers[AT]statslab.cam.ac.uk.
}  }
\title{Ending the  COVID-19 epidemic in the United Kingdom}
\begin{document}
\maketitle

\begin{abstract}
Social distancing and lockdown are the two main non-pharmaceutical interventions  being used by the UK government to contain and control the COVID-19 epidemic; these are being applied uniformly across the entire country, even though the results of \cite{ICreport} show that the impact of the infection increases sharply with age. This paper develops a variant of the workhorse $SIR$ model for epidemics along the lines of \cite{anderson1992infectious},\cite{towers2012social} \cite{singh2020age}, where the population is classified into a number of age groups. This allows us to understand the effects of age-dependent controls on the epidemic, and explore possible exit strategies.
\end{abstract}

\section{Introduction.}\label{intro}
The global COVID-19 pandemic has swept through the nations of the world with a frightening speed, and left governments struggling to cope with the situation. The initial responses have been directed towards limiting the death toll and ensuring that health services are not completely overwhelmed, as would be only too possible with an infection that can grow by a factor of one thousand in a month. As there is as yet no vaccine, no effective medication, and very imperfect understanding of the parameters of the epidemic, efforts have been directed towards containment, with decisions about return to normality being left until later. Without vaccine or effective medical treatment, the only remaining strategies would appear to be {\em either} a policy of contact tracing and quarantining, {\em or} developing herd immunity. The first of these policies appears to have been applied successfully in South Korea and Singapore, and is generally regarded as the first line of public health defence. In the current pandemic, most countries have quickly found themselves overwhelmed by the scale and speed of the outbreak, and have been unable to apply contact tracing as rigorously and universally as is needed for the method to work. When it does work, contact tracing and quarantine will allow an outbreak to be snuffed out before it spreads widely, but it will of course leave a large population of susceptibles open to a new infection, so continuing vigilance is essential.
As we have seen contact tracing overwhelmed, the goal of this paper is to explore the route to herd immunity, using age-dependent release from lockdown, and a gradual relaxation of social distancing rules.  In Section \ref{S1} we present the model, which is in almost all respects a straightforward variant of the standard $SIR$ epidemic model. The equations contain terms for the controls which are available to modify the dynamics of the epidemic.  The problem is a control problem, and for this we have to define the objective, which we do in Section \ref{S2}. The issue is of course that we have a conflict between the obvious cost of the numbers of citizens whose lives are ended prematurely, which is a concern for the next few months;  and the damage that an extended lockdown will do to the economy, which will be a concern for many years if the aftermath of the 2008 financial crash is any guide. In setting up the cost structure, some relatively arbitrary (but hopefully reasonably realistic) assumptions have to be made; these are not in any way essential to the approach, and can easily be changed by any reader prepared to play with the Jupyter notebook posted online\footnote{https://colab.research.google.com/drive/1tbB47uSGIA0WehY-hvIYgdO0mpnZU5A8}. Parameter values, or even the entire form of the costs, can be changed by anyone with a little knowledge of Python. Experts in health economics would doubtless be able to suggest values that better embody current thinking, and before any of the results of this paper can be relied on, such inputs will be necessary.

In Section \ref{S3} we briefly discuss the datasources used, and in Section \ref{S4} we present the results of computation in various scenarios.

\section{Model formulation.}\label{S1}
A simple SIR epidemic model is too crude to allow us to model and control the key features of the COVID-19 epidemic; many infected individuals are asymptomatic, and the impact of the infection on different age groups is very different.  So we will break down the population into $J$ age groups, and let $A_j(t)$, $I_j(t)$, $S_j(t)$ denote the numbers of $j$-individuals at time $t$ who are (respectively) asymptomatic infected, symptomatic infected, and susceptible. We will denote by $N_j(t)$ the total number of $j$-individuals in the population at time $t$, and allow this to change gradually with the  influx of new births, visitors from other countries; this is to model the possibility that new infecteds come in from outside and reignite the epidemic.

\medskip

The most basic form of the evolution is governed by the differential equations
\begin{eqnarray}
\dot{I_j}(t) &=& -\rho I_j(t) + p \lambda_j(t) S_j(t)
\label{Idot}
\\
\dot{A_j}(t) &=& -\rho A_j(t) + (1-p) \lambda_j(t) S_j(t) + \iota_j(t)
\label{Adot}
\\
\dot{S_j}(t) &=& -\lambda_j(t)S_j(t) + \sigma_j(t) + \varepsilon\rho\{ \,I_j(t) + A_j(t)\, \}
\label{Sdot}
\\
\dot{N_j}(t) &=& \sigma_j(t) + \iota_j(t),
\label{Ndot}
\end{eqnarray}
where $\iota_j$ and $\sigma_j$ are known functions of time representing the arrival of new asymptomatic infectives and susceptibles respectively\footnote{Any symptomatic visitors attempting to join the population are assumed to be prevented from entering.}; and the final term on the right-hand side of \eqref{Sdot} allows for the possibility that removed infectives may not in fact be immune, and some may return to the population ready for reinfection. The parameter $p \in (0,1)$ appearing in \eqref{Idot}, \eqref{Adot} is the probability that  a susceptible becoming infected is symptomatic; and the parameter $\rho>0$ is the recovery rate. The {\em infection rates} $\lambda_j(t)$ are explicit non-linear functions of the state of the system that will be discussed shortly, but, aside from the terms involving $\lambda$, the evolution is linear. So if we stack the variables into a single vector
\begin{equation}
Z_t = [I(t), A(t), S(t), N(t)]
\label{Zdef}
\end{equation}
the evolution \eqref{Idot}-\eqref{Ndot} can be written as
\begin{equation}
\dot{Z}(t) = M Z(t) + \Lambda(t) Z(t) + \eta(t),
\label{Zdot}
\end{equation}
where $M$ is a $4J \times 4J$ constant matrix, $\Lambda$ is a simple matrix whose entries involve the $\lambda_j$ in the appropriate entries, and $\eta$ is the vector of driving terms.

\bigbreak
\noindent
[{\sc Remark.}{\em  The model \eqref{Idot}-\eqref{Ndot} is the fluid limit of a Markov chain model in which $\rho$ is the rate that an individual jumps from an infected state to the removed state, and therefore the implicit (Markovian) assumption is that the time spent in the infective state is exponentially distributed. This assumption does not fit well with observation, so we can allow for different distributions by the familiar trick of the method of stages (see, for example, \cite{barbour76}), in which a infected individual passes through a number of exponentially-distributed stages. In more detail, we can suppose that there are $K_x$ stages for the symptomatic infection, and that $I_j^k(t)$ is the number of symptomatic $j$-individuals at stage $k$ of the infection at time $t$, $j=1, \ldots, J$, $k=1,\ldots,K_x$. Making this change, the equation \eqref{Idot} becomes the system
\begin{eqnarray}
\dot{I_j^1}(t) &=& - \rho K_x I^1_j(t) + p \lambda_j(t) S_j(t)
\\
\dot{I_j^k}(t) &=& -\rho K_x (I^k_j(t) - I_j^{k-1}(t)) \qquad(k=2,\ldots,K_x)
\end{eqnarray}
This corresponds to making the duration of symptomatic infection a sum of $K_x$ independent exponentials each with mean $1/\rho K_x$, which has the same mean as an exponential of rate $\rho$ but smaller variance. We could similarly decompose the asymptomatic infections, and indeed by further ramifications of the method of stages we could make the distribution of infected time approximate any desired distribution. There is a good reason not to take this too far, however; in the numerics, the differential equation has to be solved many times. It is remarkable that this can be done in a reasonable amount of time, but the more complicated the model, the slower this step becomes and ultimately the computation will be too slow.

But however we do this, when we stack all the variables into a big state vector $Z$, the evolution still has the form \eqref{Zdot}, and the appropriate form of this is coded into the Jupyter notebooks.}]

\bigbreak

\subsection{The specification of $\lambda$.}\label{lambdaspec}
Each individual spends part of the waking day at home, and part of the waking day outside\footnote{A more refined decomposition of time outside the home can be made, but this does not really change the principles being explained here. }. 
We shall denote by $m^O_{ij}$ the mean number of contacts that an $i$-individual has per day with $j$-individuals when outside the home; and by $m^H_{ij}$ the mean number of contacts that an $i$-individual has per day  with $j$-individuals when inside the home. 

 It is important to understand that $m^O_{ij}$ is the mean number of contacts that an $i$-individual has with $j$-individuals {\em if everyone spends their entire waking day} outside the home. If the $i$-individual spends a fraction $\varphi_i$ of the waking day outside the home, and $j$-individuals spend a fraction $\varphi_j$ of the waking day outside the home, then the mean number of contacts per day which an $i$-individual has with $j$-individuals outside the home will be $\varphi_i m^O_{ij} \varphi_j$.

\medskip

Each time an infected person has contact with someone, infection will be transmitted with probability $\beta$, though of course this will only result in a change if the person contacted was susceptible. Thus the overall rate at which infection is passed in the outside world  to $j$-individuals  will be
\begin{equation}
\lambda_j^O(t) S_j(t) = \beta\sum_i\{\,A_i(t) + \delta I_i(t)\,\}\; \varphi_i(t)\,
m^O_{ij} \,\varphi_j(t)\; S_j(t)/N_j(t),
\label{eq1}
\end{equation}
where $\varphi_i(t)$ is the fraction of time spent in the outside world by $i$-individuals, and $\delta \in [0,1]$ is the proportion of symptomatic infecteds who go into the outside world. In an ideal situation, this would be close to zero, but many people with the infection get only mild symptoms and may not self-isolate. The number of $j$-individuals at time $t$ is $N_j(t)$, so  the factor $S_j(t)/N_j(t)$ on the right-hand side of \eqref{eq1} is the probability that a contacted $j$-individual is susceptible.

This may have the appearance of a conventional extension of an SIR model, but one point to flag straight away is that {\em the controls $\varphi$ enter {\bf quadratically} in the expression for the infectivity}, whereas some authors use only a linear dependence. This is erroneous.

\medskip

What happens in the home is rather more difficult to deal with. We could simply take \eqref{eq1} and change superscript $O$ to $H$, and $\varphi$ to $1-\varphi$, but this would be incorrect, because an infected individual outside may go through the day and infect a large number of people, but within the home there are relatively few that could be infected, so the scope to spread infection is much reduced - this is after all the rationale for locking down populations.  

Without the constraint on the number of infections imposed by the household size, a single infected $i$-individual in the home would be firing infection transmissions at $j$-individuals at rate
\begin{equation}
\gamma_{i,j}(t) = \beta(1-\varphi_i(t)) m^H_{ij}(1-\varphi_j(t)) .
\label{ga_def}
\end{equation}
Thus if $\tau$ is the mean infective time, during the period of infectivity each infected $i$-individual in the home will fire a Poisson($\gamma_{ij}(t)\tau$) number of infections towards $j$-individuals, and therefore will fire a possible $Z \sim  {\rm Poisson}(\gamma_i(t) \tau)$ number of infections towards all others, where $\gamma_i(t) = \sum_j\gamma_{ij}(t)$. However, the number of infections that can strike another individual cannot exceed $N-1$, where $N$ is the size of the household in which the infected $i$-individual lives. Data from the Office of National Statistics allow us to deduce the distribution\footnote{The data given groups together all households with 6 or more members, so we assume that all such households have exactly 6 members.} of $N$. The mean number of individuals at whom the infected $i$-individual fires infections is
\begin{equation}
\mu_i(t) = E[ Z \wedge (N-1) ] = \sum_{k\geq 1} P(Z \geq k) P(N \geq k+1).
\end{equation}
This is the mean number of infections the infected $i$-individual could fire at others during a period of mean length $\tau$, so we will simply suppose that while infected 
\begin{center}
{\em the $i$-individual in the home will be firing infections at rate $\mu_i(t)/\tau$.}
\end{center}
An infection fired at another will be supposed to strike a $j$-individual with probability $p_{ij}^H(t)$ proportional to $m_{ij}^H (1-\varphi_j(t))$; and given that it strikes a $j$-individual, the probability that a new infection results will be $S_j(t)/N_j(t)$.
Thus the analogue of \eqref{eq1} for new infections of $j$-individuals in the home will be
\begin{equation}
\lambda^H_j(t) S_j(t) = \sum_i \{\,A_i(t) + \delta I_i(t)\,\}\; 
\frac{\mu_i(t)}{\tau} \;p_{ij}^H(t) \frac{S_j(t)}{N_j(t)}
\label{eq4}
\end{equation}
We combine these to give finally
\begin{equation}
\lambda_j(t) = \lambda_j^O(t) + \lambda_j^H(t).
\label{lambdadef}
\end{equation}

\bigbreak
\noindent
[{\sc Remark.} {\em These assumptions represent a compromise; any honest treatment of what goes on within households would appear to require a decomposition of the population into groups according to different household compositions by age, meaning the size of the statespace gets out of control - which would render the calculation impractical.} ]

\medskip
It is worth emphasizing that there are {\em just four} controlling parameters in this model: $\beta$, the probability that a contact results in a transmission; $p$, the probability that an infected person is symptomatic; $\rho$, the reciprocal of the mean infective time; $\varepsilon$, the probability that a removed infective is still susceptible. Other values which are needed for the calculations, such as the mean numbers $m^O_{ij}$, $m^H_{ij}$ of contacts, can be found from published estimates.

\section{Costs.}\label{S2}
There are three components to the cost: the cost of lockdown, the cost of social distancing, and the cost of deaths. We take them in turn.

\subsection{Lockdown costs.}\label{S31}
There will be a normal level $\bar{\varphi_j}$ for the proportion of time spent by a $j$-individual outside the home; for the purposes of the computations, the assumption here is that of the 112 waking hours of the week, 40 are spent in school or work, 20 are spent in social activities, and 52 are spent at home, making $\bar{\varphi_j}$
equal to 60/112 for all age groups. 

If a $j$-individual is locked down at level $\varphi_j(s)$ at time $s$, we propose that the cost by time $t$ should be proportional to 
\begin{equation}
\int_0^t s\, \{ \, \bar{\varphi_j}  - \varphi_j(s) \,\} \; ds.
\label{ldcost0}
\end{equation}
For constant $\varphi_j$, this will be convex in $t$, which seems realistic; a short lockdown (as for a public holiday) causes little damage, but as the time away from regular work stretches on, the damage suffered increases more rapidly, as businesses collapse and workers are made redundant. We will consider strategies where for some $0 < u < v$ (which may depend on $j$)
\begin{equation}
\varphi_j(s) = \varphi_j(0) + \frac{\bar{\varphi_j}-\varphi_j(0)}{v-u}
(s\wedge v-u)^+,
\label{ldcost1}
\end{equation}
where $\varphi_j(0)$ is the initial level of lockdown applied. At time $u$, this starts to be relaxed in a linear fashion, being fully relaxed by time $v$. Integrating \eqref{ldcost0} up to $v$ gives the cost of a $j$-individual being locked down as
\begin{equation}
C \,(\bar{\varphi_j}-\varphi_j(0) )\, \{ \;u^2/2 + u(v-u)/2 + (v-u)^2/6\;\}
\label{ldc2}
\end{equation}
for some constant $C$. If we think that the social cost of an individual being locked down for one year is $SC1$, then the constant of proportionality in \eqref{ldc2} is fixed so that the cost will be 
\begin{equation}
SC1 \times \frac{\bar{\varphi_j}-\varphi_j(0)}{\bar{\varphi_j}}\times \{ \;u^2/2 + u(v-u)/2 + (v-u)^2/6\;\} \;\frac{2}{365 \times 365}.
\end{equation}
This then has to be summed over all the members of the population, with a small reduction for retired people, who would presumably impact the economy less if they were prevented from going out.

In the numerical implementation, we fix $v = u + 5$; this reduces the number of free parameters, and in any case reflects the realistic situation that once an age group is freed from the lockdown restrictions they will quite quickly get back to normal activity.

\subsection{Social distancing costs.}\label{S32}
Social distancing imposes costs; public transport will have to run at reduced capacity, as will restaurants and theatres. But these costs are steady ongoing frictions which do not keep people away from work for months on end. If the social distancing policy means that at time $t$ the number of contacts outside the home are reduced to a fraction $SD(t) \in (0,1)$ of the normal situation, then we propose that the cost of this policy by time $t$ would be proportional to\footnote{We make the same rule for all age groups, for simplicity;  it might be hard to maintain a social distancing policy that discriminated between age groups.} 
\begin{equation}
\int_0^t  \{ \,1  - SD(s) \,\} \; ds.
\label{sdc0}
\end{equation}
The form of $SD$ is available to choose, and in the computations we suppose that $SD$ rises from the initial value $SD_0$  to final value 1 in a piecewise-linear fashion through $NSD$ stages. This allows for the possibility that social distancing could be gradually relaxed by opening more and more classes of business or public assembly. Thus at some time $u_0$,  $SD$ starts to rise to the first staged value $SD_1$ at time $v_0$, where it remains until $u_1$; from there it rises to the next staged value $SD_2$ at time $v_1$, and so on. We suppose that the levels of the stages are equally spaced, but this can easily be altered.
If there is just one stage,  the policy starts at some value $SD_0$ and at time $u$ starts to rise linearly to 1 at later time $v$, so the overall cost will be 
\begin{equation}
C (1-SD(0)) (u+v)/2.
\end{equation}
By considering the effect of social distancing for a year, we fix the constant to give cost
\begin{equation}
SC1 \times \theta_{SD} \times (1-SD(0)) \times (u+v)/730,
\label{sdc1}
\end{equation}
where $\theta_{SD} \in (0,1)$ expresses the pain of social distancing relative to lockdown. In the calculations, we will allow the profile of social distancing to be a more general piecewise-linear continuous functions, permitting social distancing to be relaxed in stages and held at intermediate values.

\subsection{Death costs.}\label{S33}
Making an estimate of the cost of the death of an individual is ethically and procedurally quite a vexed issue. For the purposes of the calculations reported in this paper and as default values used in the Jupyter notebook, the assumption is that the cost of the death of an individual is proportional to the expected number of further years that they would have lived; and that the constant of proportionality is of the same order as $SC1$, the cost of an individual being locked down for one year. So the code has a parameter {\tt deathfactor} which is used to scale $SC1$ for the calculations. 

This is only part of the story however. We need to calculate the {\em number} of deaths which will result from any particular policy, and this comes from the calculated stream of removed symptomatic infectives, coming at rate $\rho I_j(t)$ in age group $j$. Most of these will have recovered, but a percentage of these will need hospitalization, and of those a percentage will need critical care. The probabilities depend on the age of the patient, with older patients at much higher risk; estimates are given in \cite{verity} and are quoted in \cite{ICreport}.  So we calculate the rate at which new critical care beds are required. Based on an estimate for the number of days a critical care patient needs a bed (taken to be 20 days), and knowing the total number of available critical care beds, we can keep a running count of the number of critical care beds in use, and then see how many of the incoming patients for critical care can be accommodated. Those who can be accommodated survive with probability $p_{cc}$ (taken to be 0.5); those who cannot are assumed to die. It is assumed that younger patients always take priority in allocating limited resources. 

\section{Data.}\label{S3}
The code is built around the data assumptions in \cite{ICreport}, who use nine age groups, 0-9, 10-19, 20-29, 30-39, 40-49, 50-59, 60-69, 70-79, and 80+. The probabilities of hospitalization and critical care need  for these age groups are estimated by Verity {\it et al.} \cite{verity}. The population numbers for these age groups come from the Statista web site ({\tt https://www.statista.com/topics/755/uk/}). The number of critical care beds in England at the end of 2019 was around 4100, with around 11000 more planned at the emergency Nightingale hospitals, so as an optimistic figure we took 12500 to be the number. The mean infectious period was taken to be 7 days, in line with values in \cite{ICreport}, but it seems this can be highly variable. Various values were tried for $p$, the probability of an infected person being symptomatic, but the baseline for this parameter was 0.3. Infectivity  was taken to be 2.4, in line with values proposed by \cite{ICreport}, but again there appears to be quite a wide range of possible values, as we see from \cite{li2020early}. The contact matrix values $m^O$, $m^H$ are derived from \cite{prem2017projecting}; they work with different age ranges, so some pre-processing of their data had to be done; the code for this is available from the author on request.

\section{Computation.}\label{S4}
The code for the calculations was written in Python, and is available in the Jupyter notebooks for the reader to scrutinize and experiment with. The first approach was to take the objective and minimize this using the Scipy routine {\tt minimize}, which acts as a wrapper to fourteen different methods, only a few of which were possibilities due to the constrained nature of the problem. The only routine which managed acceptable runtimes was {\tt SLSQP}, but it turned out that for virtually all randomly-chosen starting points, the end point was the same as the start point; so this suggested the method which is used in the Jupyter notebook, which is simply to randomly generate control rules of the form discussed above, and focus on those which do best.

It is of course impossible to present more than just a  few cases, but we can explain what the default values for all the relevant parameters are, and then show how the outputs vary as some of these get varied. As defaults, we have taken
\begin{eqnarray*}
{\tau} &=& 7   \\
{\rm Infectivity} &=& 2.4 \\
p &=& 0.3 \\
\delta &=& 0.05\\
\varepsilon &=& 0.05 \\
{\tt phasein} &=& 5\\
K_x &=& K_y = 9\\
SD_0 &=& 0.2 \\
NSD &=& 4 \\
SC1 &=& 1e4 \\
SD_{\rm end} &=& 0.9
\\
{\tt CCstay} &=& 20\\
{\tt pcc}&=& 0.5 \\
\theta_{SD} &=& 0.25 \\
{\tt deathcosts} &=& [82., 72.22, 62.5, 52.86, 43.33,
       34.        , 25.        , 16.67, 10.        ]*SC1
\end{eqnarray*}
As initial values, we assume there are 50 asymptomatic infecteds in each of the 9 age groups, and the initial vector $\varphi_0$ is
\begin{equation*}
\varphi_0 = [ 5,5,10,10,10,10,5,5,5] * \bar{\varphi} /100
\end{equation*}
The costs of lockdown are supposed to be less severe for the older age groups, so we use 
\begin{equation*}
{\tt qcost} = [1,1,1,1,1,1,0.5,0.25,0.1]  * SC1 
\end{equation*}
As mentioned before, we took the number {\tt Nbeds} of critical care beds to be 12500. We ran the calculation for 900 days (except in the do-nothing example, which ran for 200 days). We insisted that lockdown ends for all but the oldest age group (80+) by day 400, and we imposed the condition that social distancing reaches its end value $S_{\rm end}$  by day 840.

\subsection{Base case: do nothing.}\label{S5.1}
\begin{figure}[H]
 \caption{Do nothing epidemic.}
  \centering
   \includegraphics[scale=0.35]{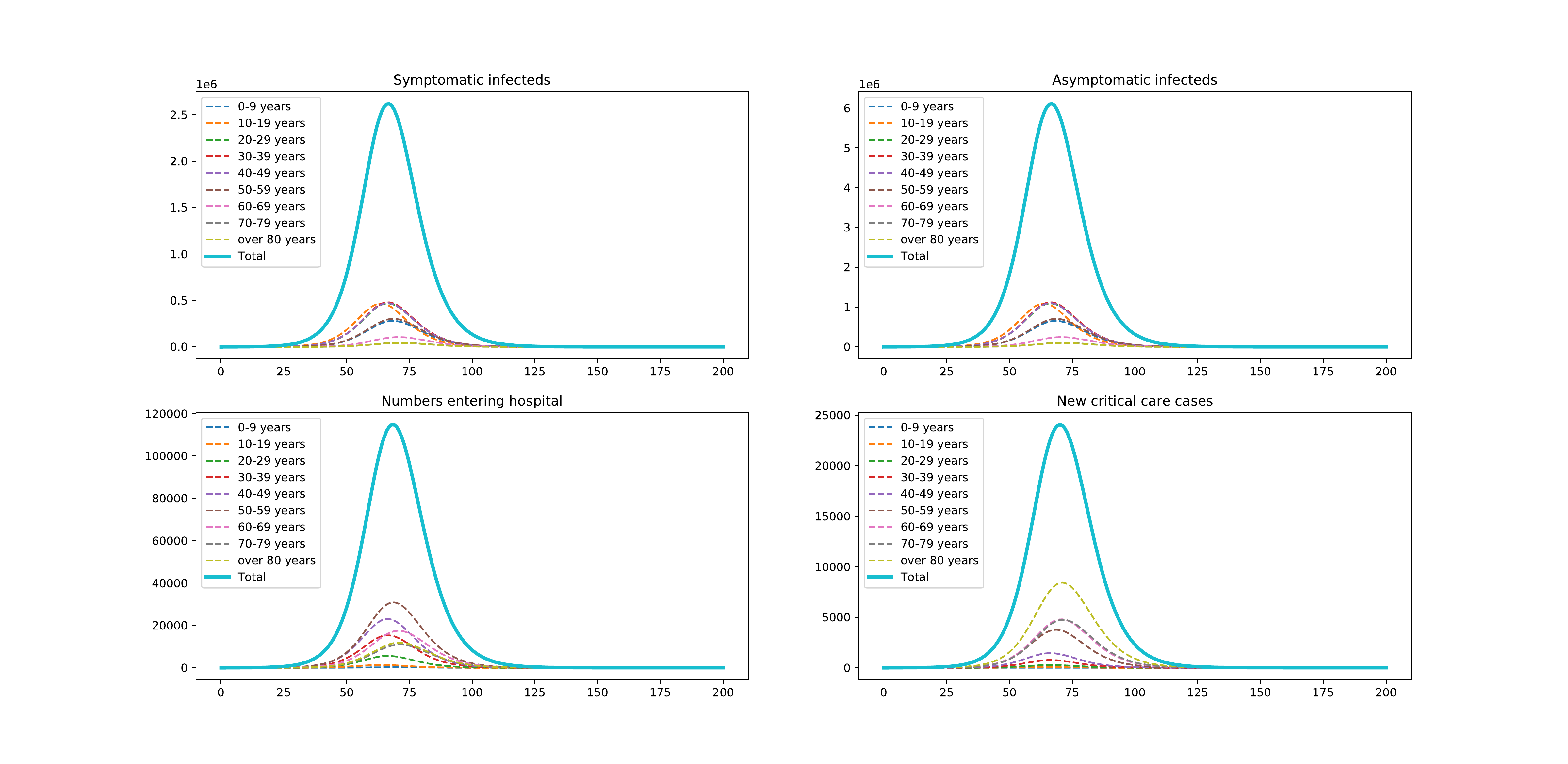} 
   \label{fig51a}
\end{figure}

\begin{figure}[H]
  \caption{Do nothing epidemic.}
  \centering
   \includegraphics[scale=0.35]{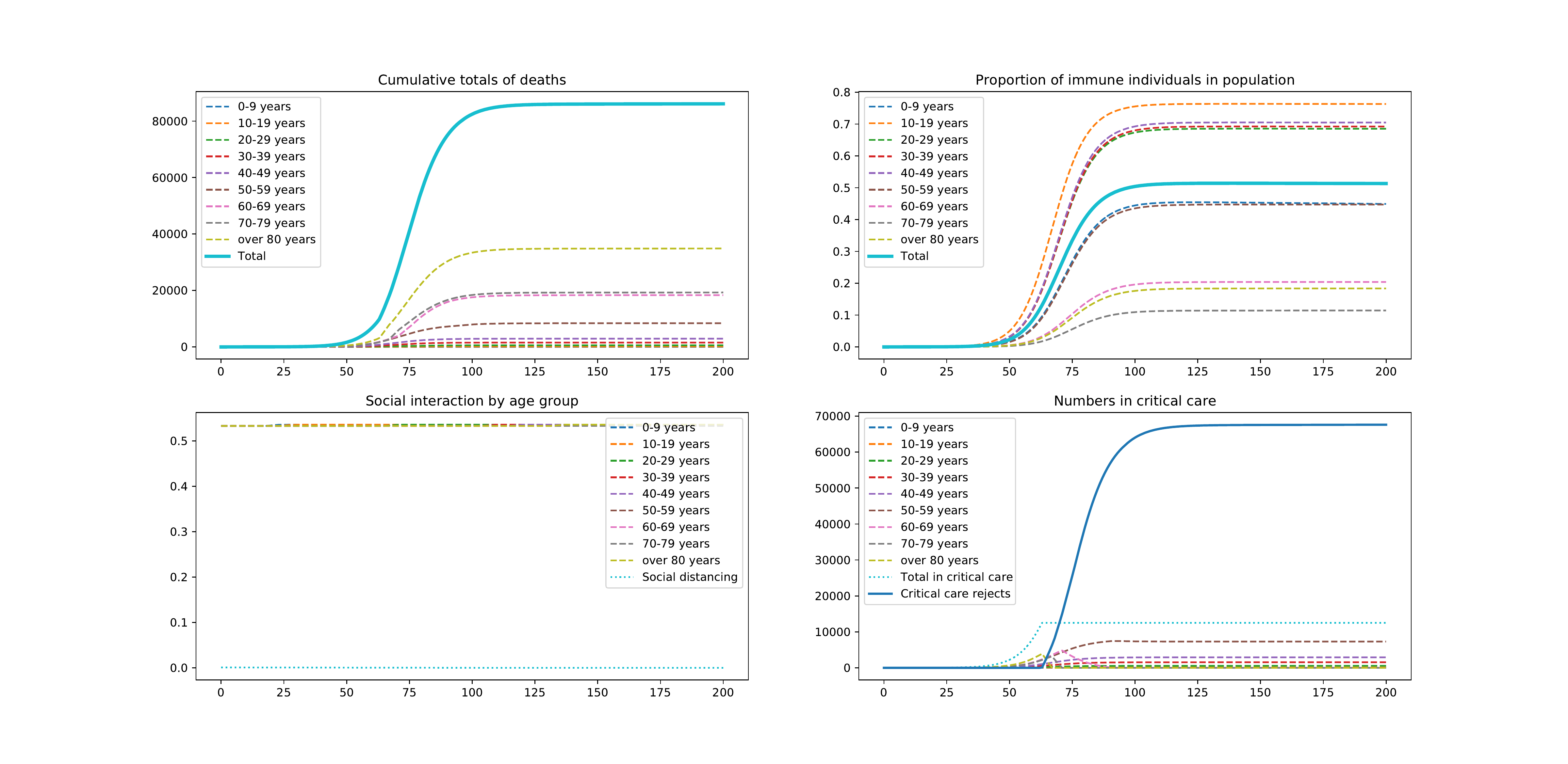} 
   \label{fig51b}
\end{figure}

In this base case, we shall take $SD_0 = 0.995$ and $\varphi_0 = \bar{\varphi}$, which is the situation where no social distancing and no lockdown happens. There are 86,000 deaths, and using the proposed cost parameters, the cost of deaths is £16bn, the cost of social disruption is £0.26bn. In this scenario, the epidemic is short and massive; as we see from Figure \ref{fig51a}, everything is over in about 100 days, with a peak number of new daily cases for critical care of 25,000, and for hospital admissions of almost 120,000. Figure \ref{fig51b} shows that the critical care provision is completely swamped, with nearly 70,000 critical care cases unable to get a critical care bed and therefore dying without the necessary care. It is hard to imagine how such a scenario could be thought acceptable.

\pagebreak
\pagebreak
\subsection{Lockdown and social distancing.}\label{S52}
In this scenario, fairly tight lockdown and social distancing measure are applied from the beginning and gradually relaxed. The costs of lockdown and social distancing this time amount to £146bn, the death costs to £4.4bn, and the total number of deaths  was 16,100. The load of new cases is much more manageable, with a peak of just over 2,200 new critical care cases, and about 16,000 new cases in all. All but the two oldest age groups are out of lockdown within 100 days, but looking at Figure \ref{fig52b} we see that even after 800 days the epidemic is far from over; once the oldest group is let out of lockdown and social distancing has come to an end, the epidemic starts to take off again. Most worrying here is that from 500 days on, every single critical care bed is taken by a COVID-19 patient, and thousands of elderly patients needing a critical care bed are unable to obtain one. This supports the proposition that some form of social distancing will have to be maintained for a very long time if no treatment or vaccine can be found.

\pagebreak
\begin{figure}[H]
  \caption{Lockdown and social distancing.}
  \centering
   \includegraphics[scale=0.35]{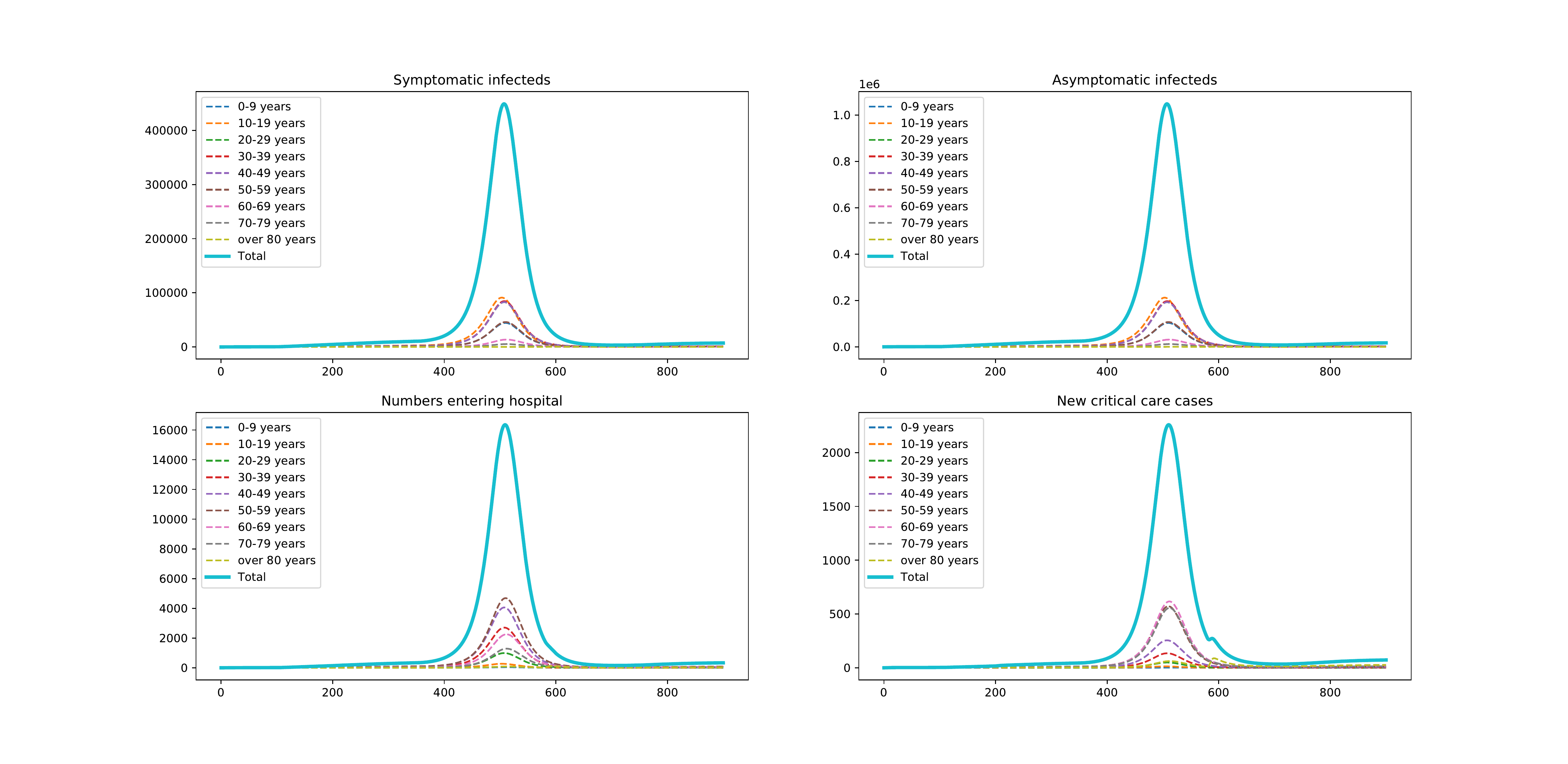} 
   \label{fig52a}
\end{figure}
\begin{figure}[H]
  \caption{Lockdown and social distancing.}
  \centering
   \includegraphics[scale=0.35]{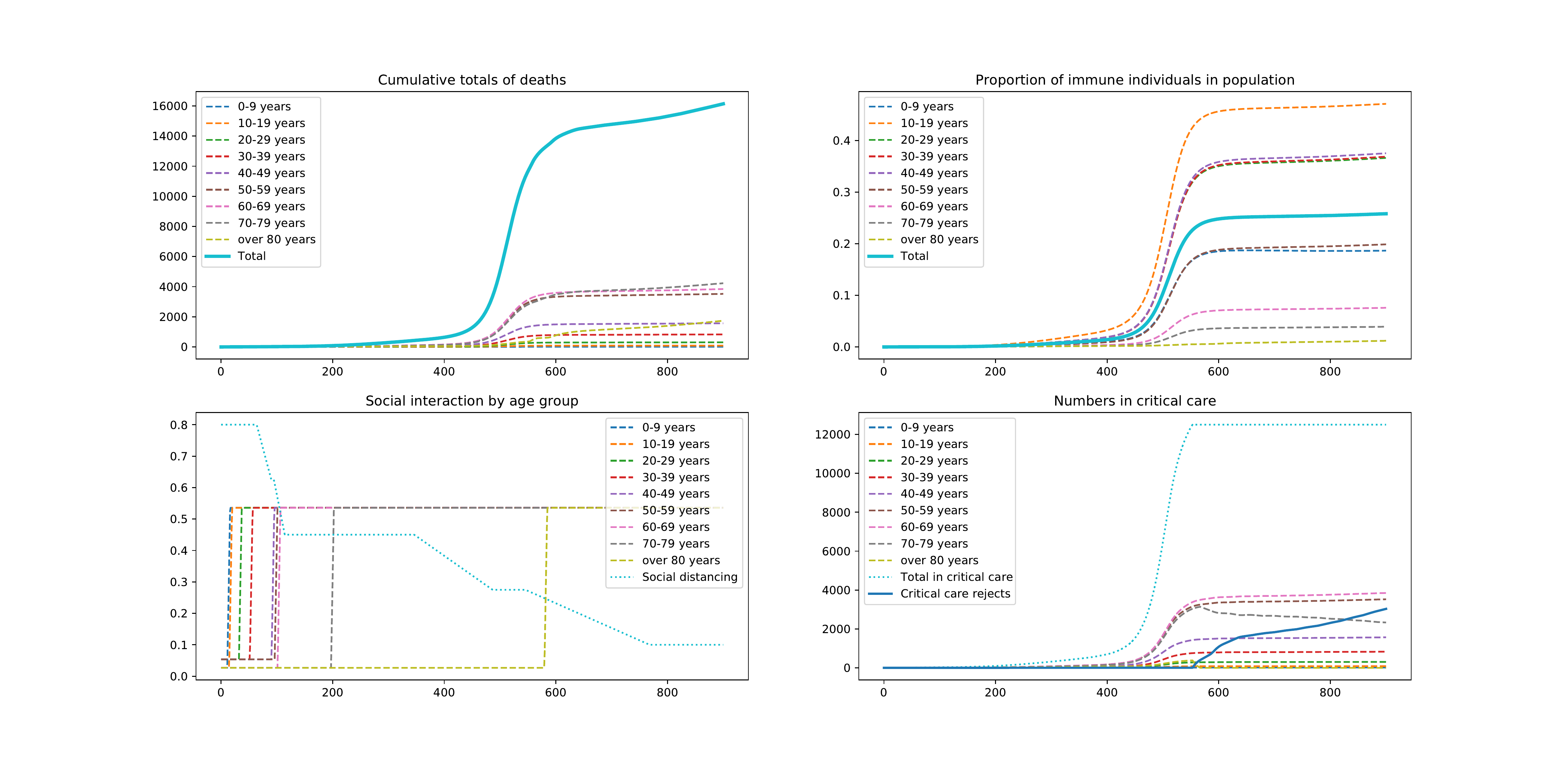} 
   \label{fig52b}
\end{figure}
\pagebreak

\subsection{Lockdown and social distancing with Infectivity = 2.8.}\label{S53}
Next we see what happens if in fact the infectivity is higher than the middle case value of 2.4 suggested in \cite{ICreport}. This time, lockdown and social distancing costs remain at around £146bn, death costs are about £8.5bn, and the total number of deaths is  38,700. The general picture looks like the previous situation but more accentuated; there is a clear second surge after the oldest age group is released from lockdown, and some 25,000 die without the critical care they need as the hospitals are submerged with cases. This time, saturation of the critical care facilities begins around day 600 and keeps going. Even maintaining social distancing at 90\% is not sufficient to hold back the epidemic in the longer run.

\pagebreak
\begin{figure}[H]
  \caption{Lockdown and social distancing, $R_0 = 2.8$}
  \centering
   \includegraphics[scale=0.35]{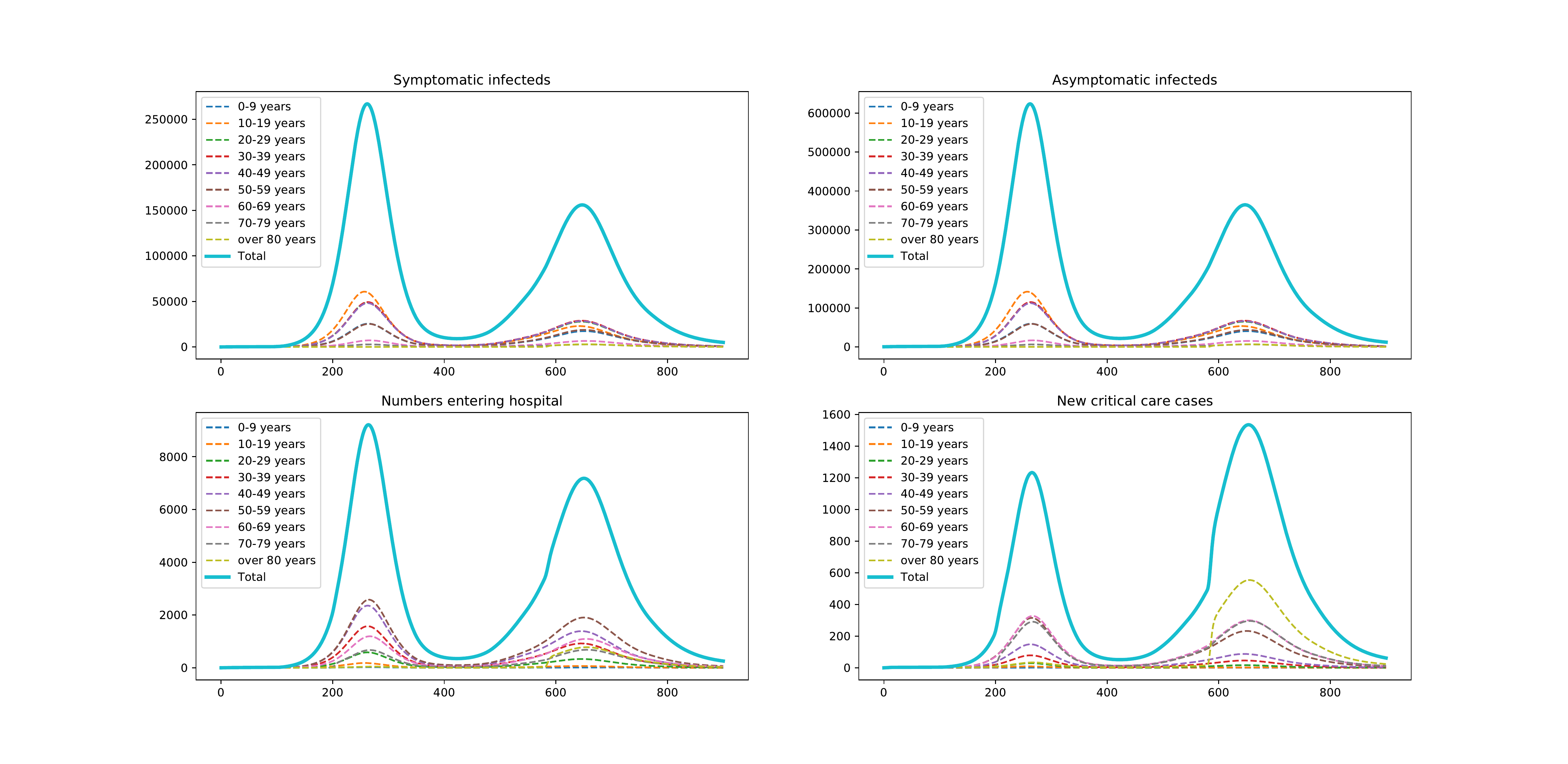} 
   \label{fig53a}
\end{figure}
\begin{figure}[H]
  \caption{Lockdown and social distancing, $R_0=2.8$.}
  \centering
   \includegraphics[scale=0.35]{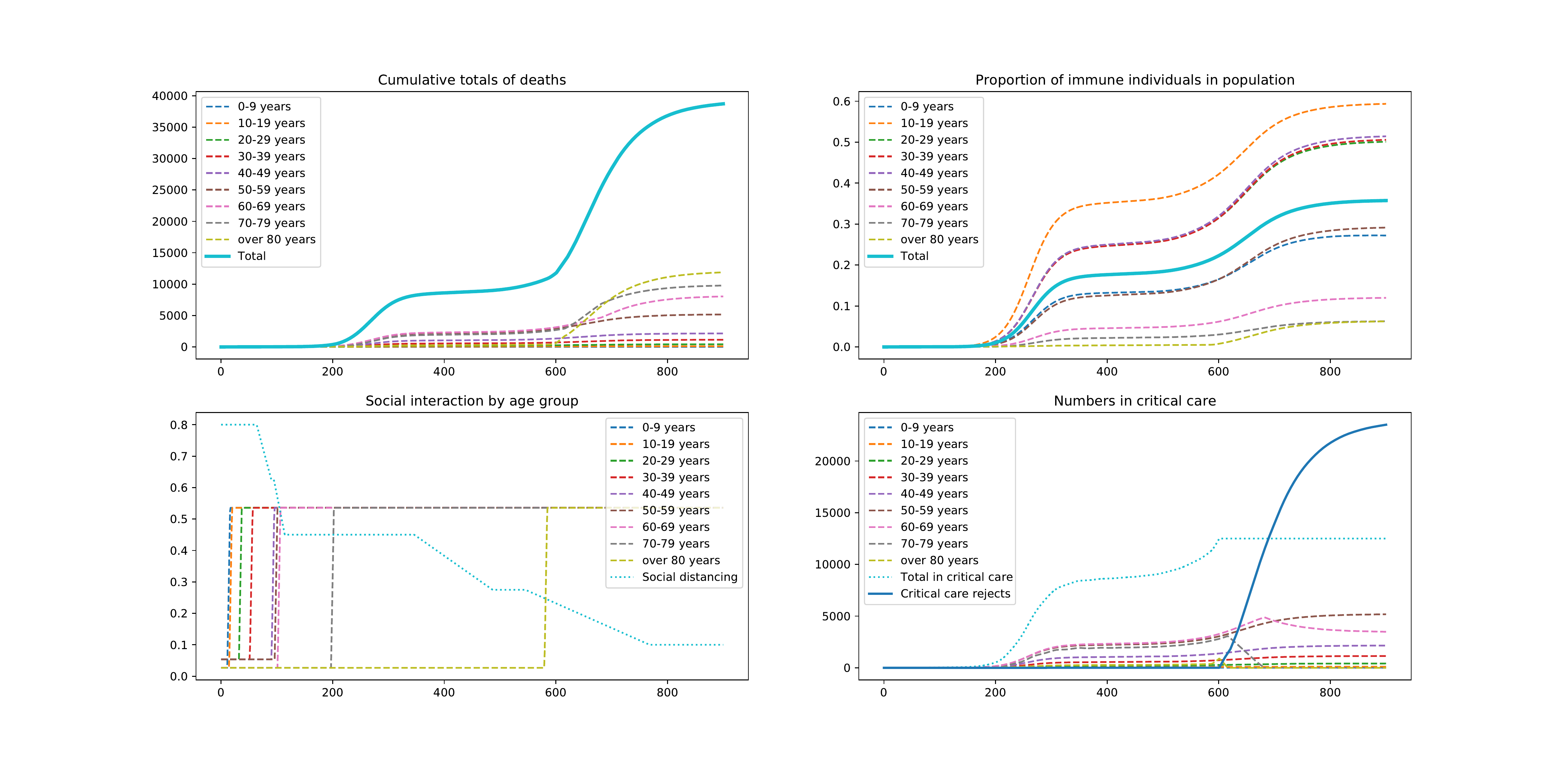} 
   \label{fig53b}
\end{figure}
\pagebreak

\subsection{Lockdown and social distancing with $p = 0.2$.}\label{S54}
\begin{figure}[H]
  \caption{Lockdown and social distancing, $p=0.2$}
  \centering
   \includegraphics[scale=0.35]{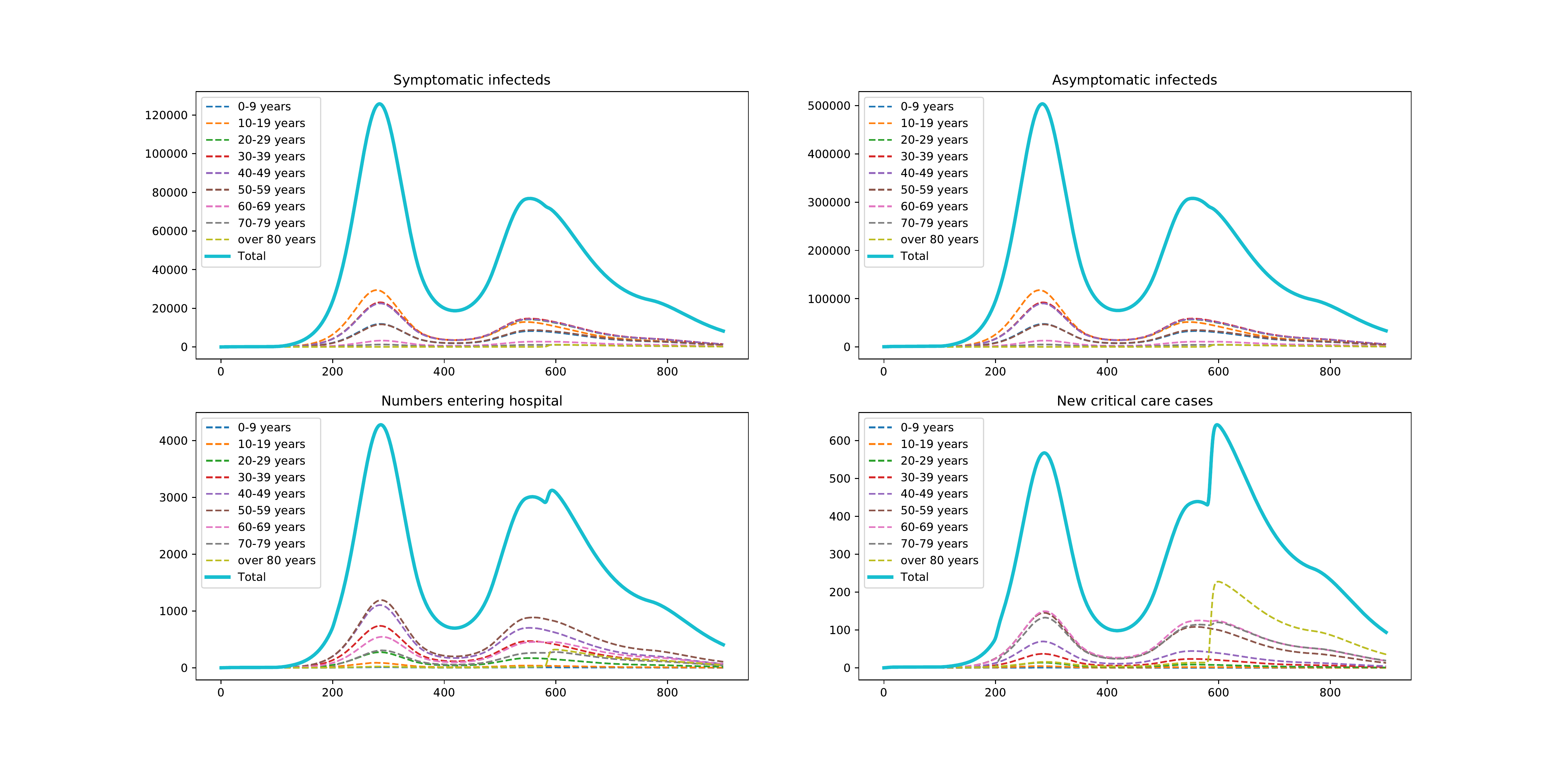} 
   \label{fig54a}
\end{figure}

If the probability that an infective is symptomatic is reduced to 0.2, the outcome is improved, with death costs around £4.2bn, lockdown costs little changed, and total deaths reduced to 17,500. Figure \ref{fig54a} shows two pronounced peaks to the infection, the second again coinciding with the final relaxation of restrictions. The critical care capacity only saturates at around day 650 this time.
The epidemic is on a smaller  and more manageable scale; peak admissions to critical care are just over 600, peak hospital admissions just over 4000. This is not surprising, since the proportion of those infected who are symptomatic (and therefore open to possible complications) is lower. However, there are more undetected asymptomatic infecteds going about in the population, so the number of deaths is higher than in the base case; it is clear from the pictures that towards the end the epidemic is beginning to get out of control. 
\pagebreak
\begin{figure}[H]
  \caption{Lockdown and social distancing, $p=0.2$.}
  \centering
   \includegraphics[scale=0.35]{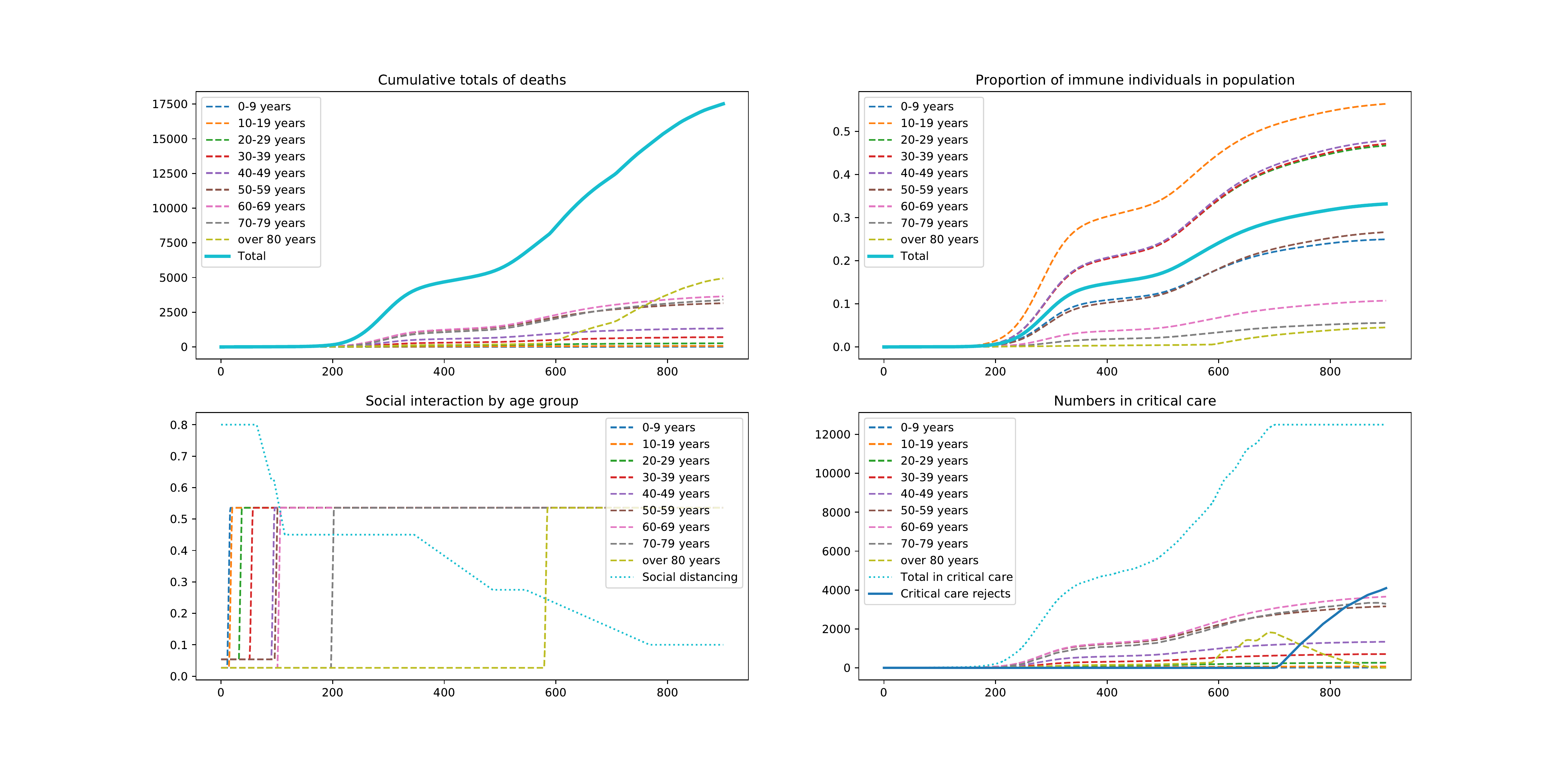} 
   \label{fig54b}
\end{figure}
\pagebreak

\subsection{Lockdown and social distancing with $SD_{\rm end} = 1.$}\label{S55}
In this scenario, we find the costs of lockdown and social distancing to be reduced to £124bn, death costs around £5.7bn. The number of deaths is 22,700. What is most clear from Figure \ref{fig55b} is that from the time that the 70-79 age group is released from lockdown around day 200, the epidemic gradually gets more out of control, with critical care at full stretch from day 450 onwards, and the numbers of older patients needing critical care and dying without it growing  10,000 at the end of the run.

\pagebreak
\begin{figure}[H]
  \caption{Lockdown and social distancing, $SD_{\rm end} = 1.$}
  \centering
   \includegraphics[scale=0.35]{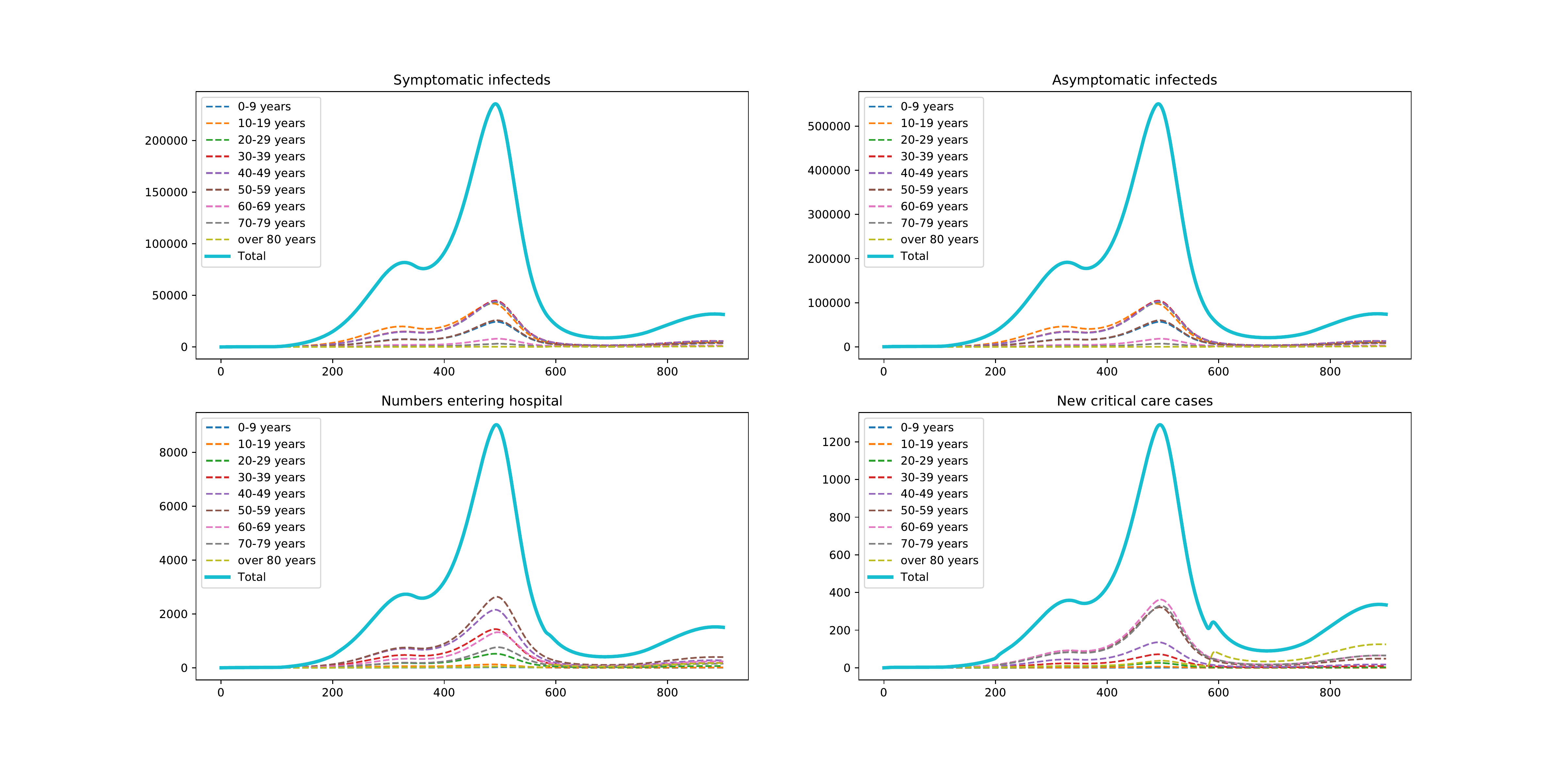} 
   \label{fig55a}
\end{figure}
\begin{figure}[H]
  \caption{Lockdown and social distancing, $SD_{\rm end} = 1.$}
    \centering
   \includegraphics[scale=0.35]{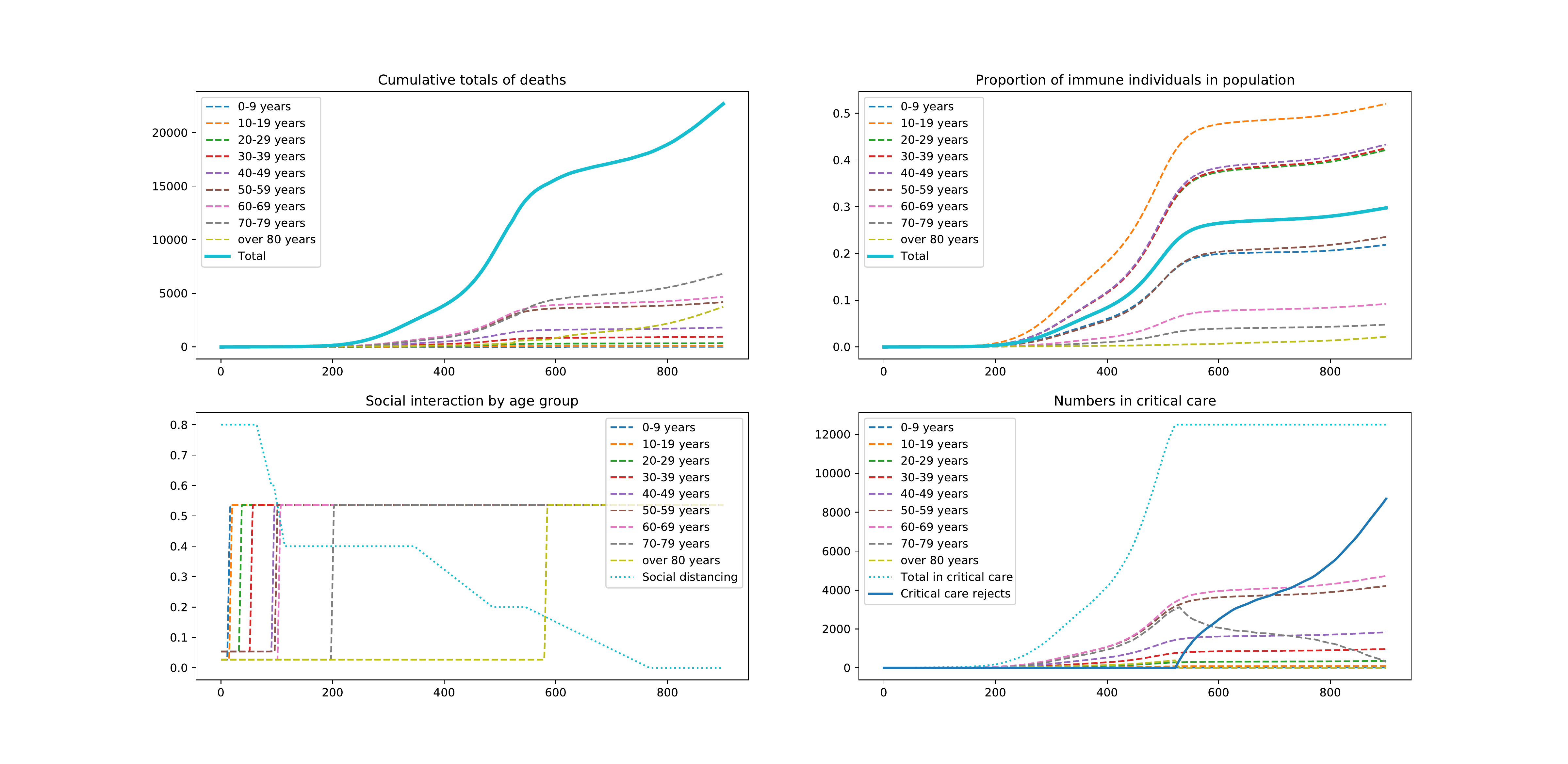} 
   \label{fig55b}
\end{figure}
\pagebreak

\bigbreak
Of course, it is only possible to display a few examples, which barely begins to explore the diversity of behaviour that will arise as parameters are varied. This is the purpose of the Jupyter notebook which can be found at
\begin{center}
https://colab.research.google.com/drive/1tbB47uSGIA0WehY-hvIYgdO0mpnZU5A8
\end{center}

\section{Conclusions.}\label{conc}
This paper offers a simple model for the current COVID-19 epidemic; no account is taken of spatial effects, which could make a big difference to any conclusions. The treatment of the spread of the infection in the home is an approximation, plausibly based perhaps, but still an approximation.  Nevertheless, the modelling assumptions are simple and compact, and permit rapid exploration of possible responses of a non-pharmaceutical nature. The calculations require assumptions about the initial state of the epidemic which are essentially guessed. Even coming into the epidemic once it is under way, it would be hard to get reliable values for the numbers of asymptomatic, susceptible and immune people in the population, not least because there is at the time of writing no test to determine whether someone has had the infection and is now immune, and only a rather unreliable test whether an individual currently has the infection.
No account is taken of parameter uncertainty. This is a natural area of enquiry, but at the moment it seems that the data that would support strong conclusions is not yet available. As it seems that the key parameters are known with very little precision, a highly detailed model, or a sophisticated story about statistical inference may help less than some rough exploration of possible parameter combinations; as the epidemic evolves around the world, we will undoubtedly learn more of its characteristics, which will allow us better to control it.

\section*{Acknowledgements.} It is a pleasure to thank Josef Teichmann, Kalvis Jansons, Ronojoy Adhikari, Rob Jack, Philip Ernst and Mike Cates for illuminating discussions. As economists will insist on noting, they are not responsible for the errors herein.

\pagebreak
\bibliography{cv}

\end{document}